# Terahertz Polaron Oscillations of Electrons Solvated in Liquid Water

Ahmed Ghalgaoui, Benjamin P. Fingerhut, Klaus Reimann, Thomas Elsaesser, and Michael Woerner[*]

*Max-Born-Institut für Nichtlineare Optik und Kurzzeitspektroskopie, 12489 Berlin, Germany*

The terahertz (THz) response of solvated electrons in liquid water is studied in nonlinear ultrafast pump-probe experiments. Free electrons with concentrations from $c_e = 4$ to $140 \times 10^{-6}$ moles/liter are generated by high-field THz or near-infrared multiphoton excitation. The time-resolved change of the dielectric function as mapped by broadband THz pulses exhibits pronounced oscillations persisting up to 30 ps. Their frequency increases with electron concentration from 0.2 to 1.5 THz. The oscillatory response is assigned to impulsively excited coherent polarons involving coupled electron and water shell motions with a frequency set by the local electric field.

A free electron solvated in liquid water, the hydrated electron, represents a prototypical quantum system that has attracted strong interest in condensed matter physics, physical chemistry, and radiology [1–3]. The quantum-mechanical cavity model describes the solvated electron as a particle locally bound in a self-consistent potential originating from a reorientation of dipolar water molecules and a concomitant minimization of the electric interaction energy [1]. This mechanism results in an equilibrium localization energy of some 3 eV and an electron ground-state wave function with a radius of approximately 250 pm. Dipole-allowed optical transitions arise between the $s$-like ground state and $p$-like excited states, resulting in a strong electronic absorption band with a maximum at 1.8 eV and a spectral width of 0.85 eV [3,4].

Both equilibrium and nonequilibrium dynamics of hydrated electrons have been studied in quantum-mechanical [5] and hybrid quantum-mechanical/molecular-mechanical calculations [1–3] and in ultrafast optical experiments [6–11]. In the experiments, free electrons have been generated by multiphoton excitation of water with a femtosecond pump pulse. The resulting absorption change has been mapped by probe pulses in an energy range from 5 meV to several eV. Theory and experiments consistently establish a period of electron localization of approximately 1 ps, during which the initially free electron [Fig. 1(a)] generates a binding potential of increasing depth and shrinking diameter by reorientation of water molecules [Fig. 1(b)]. This process is connected with a transfer of excess energy to the water environment, i.e., the electron cools down on a timescale extending to some 5 ps and eventually localizes in its ground state.

Most recently, a new mechanism of electron generation has been discovered [12]. A strong THz electric field induces a spatial separation of electrons, generated by spontaneous tunneling ionization of water molecules, from their parent ions, followed by electron localization at a new site in the liquid. This scenario has been mapped via the related changes of the THz dielectric function.

At a temperature of 300 K, the polarizable water environment displays structural fluctuations on a femto- to picosecond timescale, directly affecting the electron potential [13,14]. The relevant intermolecular degrees of freedom have frequencies between 0.1 and 25 THz (energies from 0.4 to 100 meV) [15]. Because of the strong Coulomb interaction between the electron and its solvating water shell, one expects a coupling of electronic and nuclear degrees of freedom, which may result in a polaronic response of an electron dressed by longitudinal excitations of the environment [16–18]. So far, such properties and their manifestation in the dielectric function of solvated electrons have remained unexplored. Transient THz spectra of hydrated electrons should give insight in this complex many-body scenario and the dynamic behavior of the coupled system.

In this Letter, we present new insight in the transient THz response of electrons solvated in water. Temporally and spectrally resolved pump-probe experiments reveal for the first time underdamped oscillatory changes of the THz dielectric function, which reflect a coherent polaronic response of hydrated electrons. The oscillation frequency is shown to scale with electron concentration and to correspond to a zero in the real part of the dielectric function. The oscillations originate from an impulsive excitation of the coupled system via the initial ultrafast electron relaxation.

In our experiments, the sample is a 50 $\mu$m-thick gravity-driven jet of liquid water at ambient temperature. Hydrated electrons are generated with concentrations $c_e < 10\ \mu M$ (1 $\mu M = 10^{-6}$ moles/liter) by nonlinear excitation with a THz transient centered at 0.7 THz with a peak electric field on the order of 500 kV/cm. Electron concentrations higher than 10 $\mu M$ are produced by multiphoton ionization of water molecules using an 800-nm pulse of 50 fs duration and a peak intensity between 5 and 10 TW/cm$^2$. The sample is probed by weak THz pulses transmitted through the jet and detected in amplitude and phase by free-space electro-optic sampling. The fully phase-resolved pump-probe signal is given by $E_{NL}(t,\tau) = E_{pr}^{pumped}(t,\tau) - E_{pr}(t)$. Here, $E_{pr}^{pumped}(t,\tau)$, depending on the pump-probe delay $\tau$ and the real time $t$, is the probe field transmitted through the excited sample, while $E_{pr}(t)$ represents the probe field transmitted without excitation. The real time $t$ is the time coordinate along which the THz probe field is measured [cf. Figs. 1(c), 1(d), and 1(f)]. The pump beam is chopped at 500 Hz (half the repetition rate of the laser system) to record data with and without excitation back to back, increasing the signal-to-noise ratio.

The electron concentration $c_e$ generated by the THz or optical pump pulses was determined in independent measurements with 800 nm probe pulses. The latter probe the transient electronic absorption of the generated solvated electrons. The electron concentration $c_e$ was derived from the measured absorbance change, the molar extinction coefficient at 800 nm [4], and the sample thickness.

Results for an electron concentration $c_e = 16\ \mu M$ generated by 800 nm excitation are summarized in Fig. 1. In Fig. 1(d), the nonlinear signal field $E_{NL}(t,\tau)$ is plotted as a function of $t$ (abscissa) and $\tau$ (ordinate). A cut of this signal taken at $t = 0$ [Fig. 1(e), red solid line] displays pronounced oscillations superimposed on a steplike transmission increase. For noise reduction, this transient was derived by Fourier transforming the raw data to the frequency domain, convoluting them with a Gaussian filter function centered at ($\nu_t = 0$, $\nu_\tau = 0$) with a FWHM of 4 THz and transforming back to the time domain [compare the dots and lines in Figs. 1(e) and 2(b)–2(e)]. Details of this procedure have been presented in Ref. [19]. The oscillatory part $E_{\text{osc}}(0,\tau)$ of the signal (blue line) is separated by subtraction of the steplike response. It displays a weak damping up to the longest delay time of 24 ps. The Fourier transform of this transient shows a sharp peak at $\nu_\tau = 0.5$ THz [blue line in Fig. 1(g)].

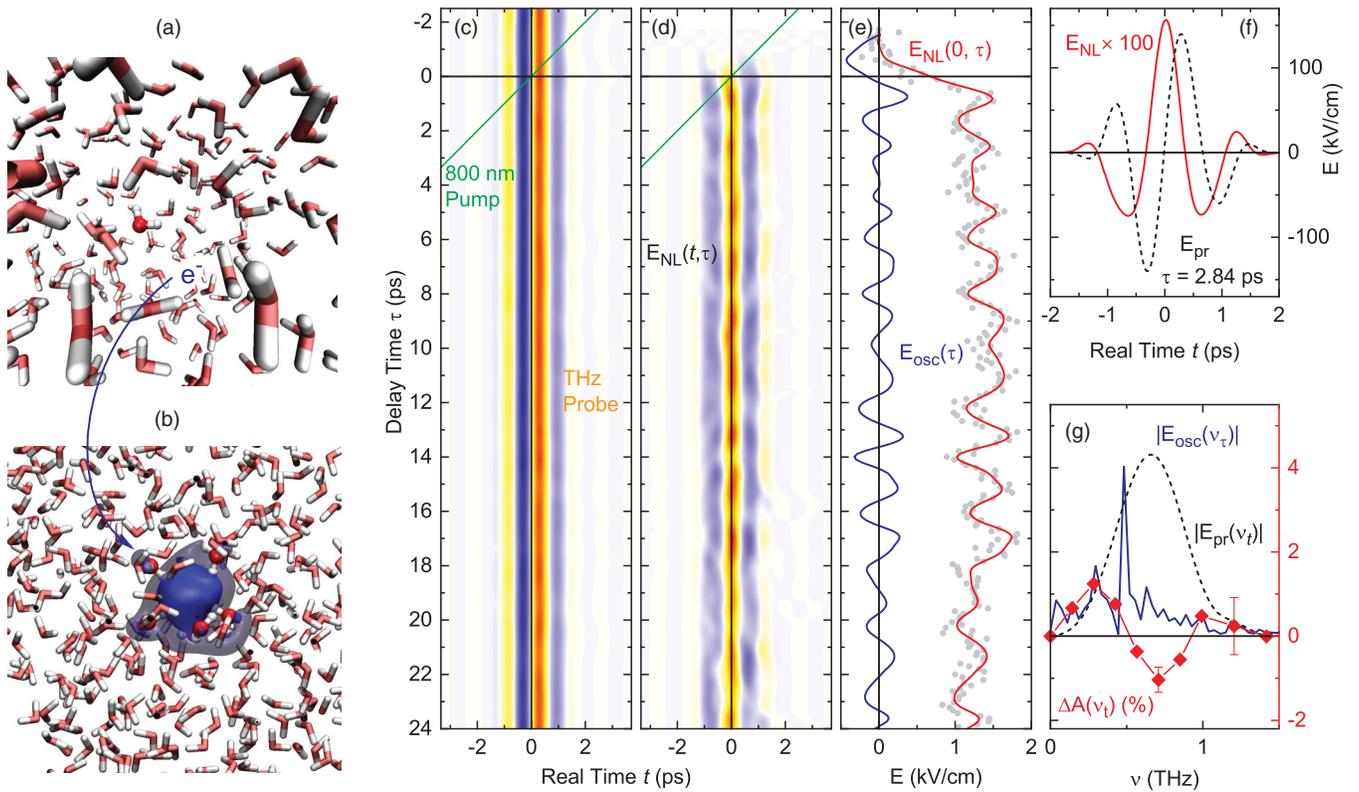

FIG. 1. Nonlinear THz response of electrons solvated in water (electron concentration of $c_e = 16\ \mu M$). (a) Molecular dynamics results showing the liquid-water environment at the instant of time a free electron is generated by multiphoton ionization and (b) after solvation in the water environment. The blue contours in panel (b) represent the singly occupied molecular orbital [iso contour values 0.03 (dark blue) and 0.015 (light blue)] of the solvated electron in equilibrium. (c) Two-dimensional scan along real time $t$ and pump-probe delay $\tau$ of the THz probe pulse $E_{pr}(t)$ transmitted through the unexcited sample, and (d) the nonlinear signal field $E_{NL}(t,\tau)$. Diagonal green lines in (c) and (d) indicate the temporal position of the pump pulse. (e) Red line: Nonlinear signal $E_{NL}(0,\tau)$ as a function of $\tau$ for $t = 0$. Dots: Data before 2D Fourier filtering. Blue line: Oscillatory signal component $E_{osc}(0,\tau)$ after subtraction of the steplike contribution. (f) Probe field $E_{pr}(t)$ (dashed line) and nonlinear signal $E_{NL}(t,\tau = 2.84\ ps)$ (solid line). (g) Dashed line: probe pulse spectrum $|E_{pr}(\nu_t)|$. Symbols: Spectrally resolved pump-probe signal $\Delta A(\nu_t)$ at $\tau = 2.84\ ps$ (symbols) and Fourier transform $|E_{osc}(\nu_\tau)|$ of the coherent signal of panel (e) (blue line).

In Fig. 1(f), the nonlinear signal $E_{\text{NL}}(t,\tau)$ at $\tau = 2.84$ ps (solid line) is plotted as a function of real time $t$, together with the probe field $E_{\text{pr}}(t)$ (dashed line) transmitted through the unexcited sample. One observes a phase shift of the nonlinear signal to earlier times $t$, a hallmark of a pronounced decrease in the real part of the THz dielectric function (and of the refractive index). A two-dimensional Fourier transform of $E_{\text{NL}}(t,\tau)$ along $t$ and $\tau$ gives the frequency-domain signal $E_{\text{NL}}(\nu_t, \nu_\tau)$ as a function of detection (probe) frequency $\nu_t$ and excitation frequency $\nu_\tau$. Integration of $E_{\text{NL}}(\nu_t, \nu_\tau)$ along $\nu_\tau$ provides the spectrally resolved pump-probe signal, from which the absorption change $\Delta A(\nu_t) = -\ln(T/T_0)$ is derived ($T$, $T_0$: intensity transmission of the sample with and without excitation). This spectrally resolved absorption change $\Delta A(\nu_t)$ shown in Fig. 1(g) (symbols) exhibits a transition from induced absorption ($\nu_t < 0.5$ THz) to bleaching ($\nu_t > 0.5$ THz). As a consequence the spectrally integrated pump-probe signal almost vanishes.

For an in-depth characterization of the oscillatory nonlinear signal, we performed measurements over a wide range of electron concentrations $c_e$. Figure 2(a) displays the 2D-THz spectrum $|E_{\text{NL}}(\nu_t,\nu_\tau)|$ for $c_e = 140\ \mu\text{M}$. The signal components between $\nu_\tau = -1$ and $\nu_\tau = +1$ THz are responsible for the steplike response, the components at $\nu_\tau = \pm 1.5$ THz for the oscillatory signal. In panels (b)–(e), oscillatory signals $E_{\text{osc}}(\tau)$ are presented for electron concentrations from $c_e = 4$ to $c_e = 140\ \mu\text{M}$, together with their Fourier transforms in panels (f)–(i). The measurement at $c_e = 4\ \mu\text{M}$ was performed with excitation by a strong THz pulse, while all other measurements were done with femtosecond 800 nm excitation. All time-resolved transients display oscillations persisting up to tens of picoseconds, i.e., the particular electron generation mechanism does not affect the character of the nonlinear response. However, the oscillation frequency increases significantly with increasing $c_e$ from 0.23 to 1.5 THz [Fig. 2(j)].

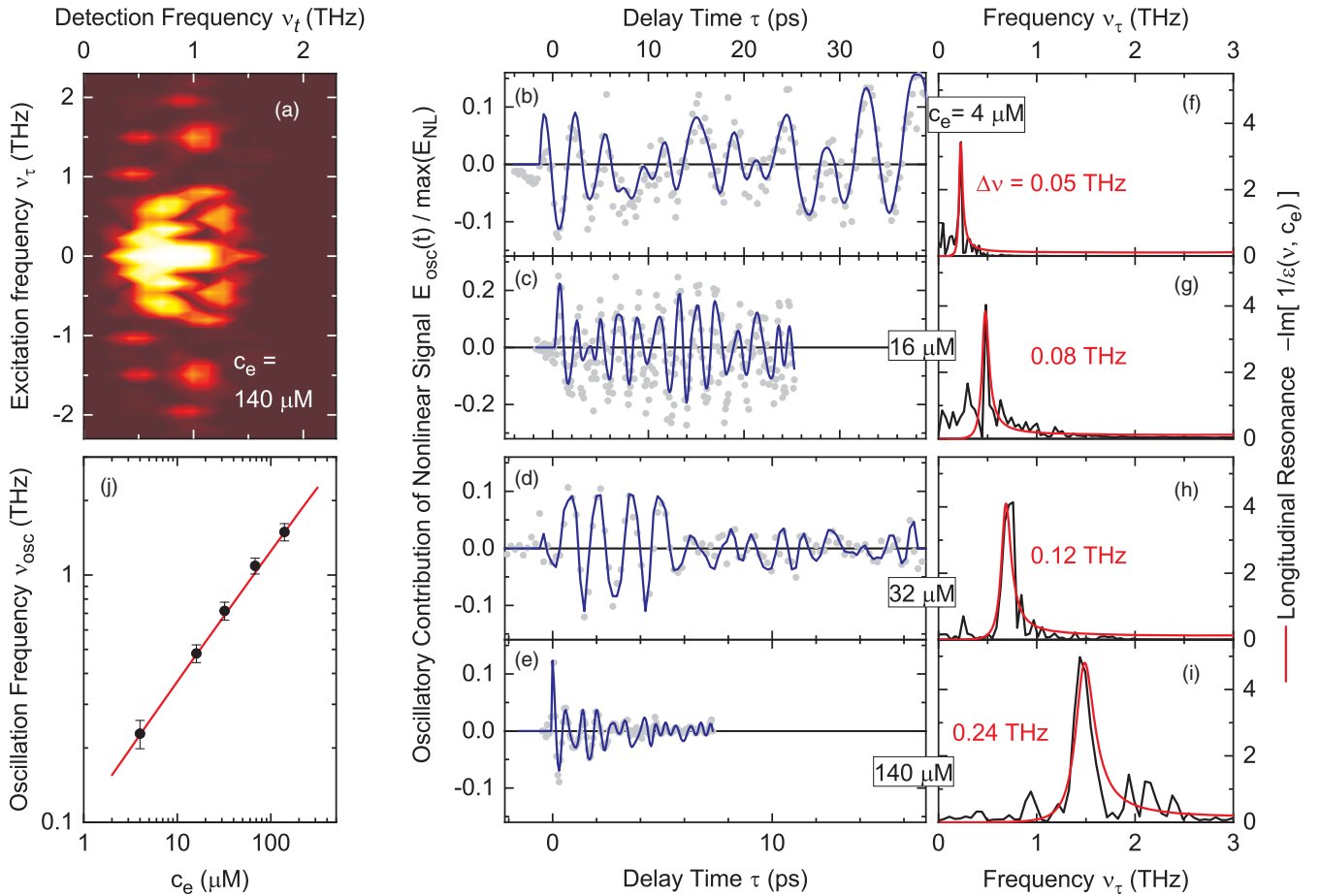

FIG. 2. (a) 2D-THz spectrum for a concentration $c_e = 140\ \mu\text{M}$ of solvated electrons. (b)–(e) Oscillatory contributions $E_{\text{osc}}(\tau)$ as a function of pump-probe delay $\tau$ for several values of $c_e$, dots before and lines after 2D Fourier filtering. (f)–(i) Black curves: Fourier transforms of the transients shown in panels (b)–(e). Red curves: Spectra $-\text{Im}[1/\varepsilon(\nu,c_e)]$ calculated from Eq. (1) with $\gamma = 0$ together with their spectral widths $\Delta\nu$ (FWHM). (j) Oscillation frequency as a function of $c_e$.

We first discuss the excitation mechanism of the pronounced coherent oscillations observed in the pump-probe signals. Ionization of a water molecule generates a quasifree electron, which initially represents a wave packet made of continuum states. After its birth, this electron undergoes a relaxation process into a spatially separated localized quantum state within 1 ps. For electron concentrations of $c_e = 4$ and 140 $\mu$M, the average distances between solvated electrons are 75 and 22 nm, respectively. Such distances are larger than the screening length of the electron Coulomb potential, which has values of 38 nm for $c_e = 4$ $\mu$M and 6 nm for $c_e = 140$ $\mu$M, assuming a THz dielectric constant of $\varepsilon_{\text{real}} = 5$. As a result, the electrostatic interaction between neighboring solvated electrons is weak compared to the local Coulomb interactions between the electron and its water shell.

The relaxation process involves real-space transport of the electron wave packet away from the ionization site, decoherence, solvation by reorientation of water molecules, and a concomitant transfer of excess energy into the aqueous environment. Both solvation and energy transfer are connected with the excitation of low-frequency vibrational and librational modes of the water environment. The (sub)picosecond relaxation period is comparable to or even shorter than the period of the relevant low-frequency modes. As a result, the latter are excited impulsively, i.e., a nonstationary superposition of quantum states is generated, which leads to molecular motions along the excited degrees of freedom. The dipolar character of the water molecules involved results in a strong electric coupling to the highly polarizable electron, giving this excitation the character of a polaron and mapping it onto the dielectric function of the coupled system. At the molecular level, the polaron excitation involves both nuclear and electron motions [18]. In our experiments, the resulting changes of the equilibrium dielectric function are monitored by the THz probe pulse.

To account for the observed frequency shift of the polaron oscillations with electron concentration $c_e$, we analyze the experimental results with a model for the dielectric function, based on the Clausius-Mossotti relation [20]. As shown in Ref. [20], local-field corrections due to dipole-dipole interactions occur between *all localized dipoles* in condensed matter, independent of the nature and resonance frequencies of the participating dipoles. For instance, the orientational dipole of individual water molecules interacts with the dipole of the strongly polarizable localized hydrated electrons. Adding electrons with a polarizability $\alpha_{\text{el}}(\nu)$ to neat water [$\varepsilon_{\text{water}}(\nu)$] leads to the following concentration-dependent dielectric function $\varepsilon(\nu, c_e)$:

$$3\frac{\varepsilon(\nu, c_e) - 1}{\varepsilon(\nu, c_e) + 2} = 3\frac{\varepsilon_{\text{water}}(\nu) - 1}{\varepsilon_{\text{water}}(\nu) + 2} + c_e N_A \alpha_{\text{el}}(\nu),$$
$$\text{with} \quad \alpha_{\text{el}}(\nu) = -\frac{e^2}{\varepsilon_0 m[(2\pi\nu)^2 + i\gamma(2\pi\nu)]}, \quad (1)$$

the Avogadro constant $N_A$, elementary charge $e$, vacuum permittivity $\varepsilon_0$, electron mass $m$, and local friction rate $\gamma$. The data for $\varepsilon_{\text{water}}(\nu)$ are taken from Ref. [15]. Resonances due to longitudinal excitations occur at the zeros of the real part of the concentration-dependent dielectric function, i.e., for $\text{Re}[\varepsilon(\nu, c_e)] = 0$. The resonances calculated with $\gamma = 0$ are shown in Fig. 2(j) as the solid line together with the experimental frequencies (symbols) as a function of $c_e$. In all cases the frequency positions are in excellent agreement with the calculated values.

The line shape of such resonances is determined by the imaginary part of the reciprocal of the dielectric function $-\text{Im}[1/\varepsilon(\nu, c_e)]$. The red lines in Figs. 2(f)–2(i) show calculated spectra using Eq. (1) for the given electron concentrations $c_e$ with $\gamma = 0$. As can be seen, also the calculated spectra agree very well with the experimental spectra (black lines). The agreement of the frequency positions demonstrates that the frequency shift of the polaronic excitations is governed by electron-concentration dependent changes of the local electric field acting on the highly polarizable solvated electrons [for $\nu = 1.5$ THz and $\gamma = 0$, the electron polarizability is $(33 \text{ nm})^3$].

The oscillatory pump-probe signals $E_{\text{osc}}$ persist for surprisingly long times [> 30 ps in Fig. 2(b)], leading to very narrow spectral widths, see Figs. 2(f)–2(i). The observed narrow spectral features show that the damping of these excitations is much weaker than the damping of transverse excitations [12], e.g., those related to electron transport in the liquid, which would result in subpicosecond damping times. The weak damping observed here points to a longitudinal character of the underlying excitations with a weak coupling to fluctuating forces from the environment.

In a system of charged or polar particles, the timescale of the dephasing of longitudinal excitations is set by the correlation function of charge density, while the dephasing of transverse excitations is governed by the correlation function of electric current [21–24]. The latter is much more susceptible to fluctuations and scattering processes than the correlation function of charge density. For instance, elastic scattering may decrease the electric current, but does not change the charge density. Accordingly, transverse excitations are more strongly damped than longitudinal excitations.

In our case, polaron damping does not originate from local particle collisions of solvated electrons, but is dominated by dielectric losses due to the orientational, i.e., Debye relaxation of water molecules. This source of damping is accounted for by the dielectric function of neat water $\varepsilon_{\text{water}}(\nu)$ in Eq. (1) and, thus, is present even for $\gamma = 0$. A related scenario has recently been analyzed for plasmas [25], where it is shown that the damping rate of plasma oscillations is much weaker than expected from collisional relaxation processes. While our results do not allow us to unambiguously identify the microscopic character of the observed oscillations, a possible scenario is size oscillations of the polaron after its generation. Such oscillations would be damped only very weakly since they do not couple directly to a macroscopic field. A detailed theoretical analysis would require a calculation with dipoles of finite size along the lines of Ref. [26].

In conclusion, our work demonstrates a novel type of oscillatory THz response of electrons solvated in liquid water. Impulsive excitation of coherent polarons leads to a strong periodic modulation of the THz dielectric function, which persists for up to tens of picoseconds. The weak damping of the oscillatory response points to a longitudinal character of the underlying excitation, coupling only weakly to fluctuating forces from the solvated electron's environment. The long decoherence times may allow for controlling dielectric properties of this system by tailored optical or THz pulses.

This research has received funding from the European Research Council (ERC) under the European Unions Horizon 2020 Research and Innovation Program (Grant Agreements No. 833365 and No. 802817). M. W. acknowledges support by the Deutsche Forschungsgemeinschaft (Grant No. WO 558/14-1).